\documentclass[prper,aps,twocolumn,groupedaddress,floats,showpacs,final,superscriptaddress]{revtex4-1}
\usepackage[latin3]{inputenc}
\usepackage[makeroom]{cancel}
\usepackage{graphicx}
\usepackage{amsmath}
\usepackage{amsfonts}
\usepackage{amssymb}
\usepackage{chngcntr}
\usepackage{color}
\usepackage[left=2cm,right=2cm,top=2cm,bottom=2cm]{geometry}

\usepackage{graphicx} 
\usepackage{dcolumn}
\usepackage{bm}
\usepackage{simplewick}
\usepackage{array}
\usepackage{appendix}

\RequirePackage[
   hyperindex,colorlinks,bookmarksnumbered,
   plainpages=true,pdfstartview=FitH]{hyperref}
\hypersetup{linkcolor=blue,urlcolor=blue,citecolor=blue}
\usepackage{hyperref}

\definecolor{purple}{rgb}{0.5,0,0.6}

\usepackage{ulem}

\begin{document}

\date{\today}
\title{Kondo effect in a Aharonov-Casher interferometer}
%\title{Interplay between Kondo effect and spin-orbit interaction in quantum nano-wires}
\author{A. V. Parafilo}
\affiliation{Center for Theoretical Physics of Complex Systems, Institute for Basic Science, Expo-ro, 55, Yuseong-gu, Daejeon 34126, Republic of Korea}

\author{L. Y. Gorelik}
\affiliation{Department of %Applied
Physics, Chalmers University of
Technology, SE-412 96 G{\" o}teborg, Sweden}

\author{M. N. Kiselev}
\affiliation{The Abdus Salam International Centre for Theoretical
Physics, Strada Costiera 11, I-34151 Trieste, Italy}
\author{H. C. Park}
\affiliation{Center for Theoretical Physics of Complex Systems, Institute for Basic Science, Expo-ro, 55, Yuseong-gu, Daejeon 34126, Republic of Korea}

\author{R. I. Shekhter}
\affiliation{Department of Physics, University of Gothenburg, SE-412
96 G{\" o}teborg, Sweden}

\date{\today}
\pacs{}

\begin{abstract}
 We consider a model describing a spin field-effect transistor based on 
a quantum nanowire with a tunable spin-orbit interaction embedded between two ferromagnetic leads with anticollinear magnetization. We investigate a regime of strong interplay between resonance Kondo scattering and interference associated with the Aharonov-Casher effect.
Using the Keldysh technique at a weak coupling regime we calculate perturbatively 
the charge current.
It is predicted that the effects of the spin-orbit interaction result in a non-vanishing current for any spin polarization of the leads including the case of fully polarized anti-collinear contacts. 
We analyze the influence of the Aharonov-Casher phase and degree of spin polarization in the leads onto a Kondo temperature.
\end{abstract}
\maketitle

\section{ Introduction} Kondo effect is known to play a very important role for a charge transport through nano-structures facilitating the maximal conductance of a nano-device at zero bias [\onlinecite{revival}]. Having a spin nature, Kondo effect is associated with a resonance scattering accompanied by the spin flip through the multiple cotunneling processes in Coulomb blockaded nano-devices [\onlinecite{rev}]. Kondo effect in GaAs based semiconductor nano-structures (quantum dots, quantum point contacts, quantum wires etc) attracted enormous attention of both experimental and theoretical communities during last two decades [\onlinecite{rev}-\onlinecite{leo}]. Recently, semiconductor quantum wires fabricated on InAs and InSb heterostructures started to be widely used in the new quantum technological devices [\onlinecite{in1}],[\onlinecite{in2}]. One of the most important property of these materials is related to effects of a strong spin-orbit interaction (SOI) which does not conserve spin in the resonance scattering processes, see, e.g., [\onlinecite{rashbareview}]. The high tunability of the interplay between SOI and resonance Kondo effect and its influence on the charge and spin transport through nano-structures paves a way for practical applications of these materials in spintronics devices. It is known that contrast to effects of external magnetic field, the effect of SOI on electrons scattering is preserving a time-reversal symmetry. 
While magnetic field is destructive for the Kondo effect due to suppression of the spin-flip processes, the influence of SOI on the resonance scattering is more delicate.

%%%%%%%%%%%%%%%%%%%%%%%%%%%%%%%%%%%%%%%%%%%%%%%%%%%%%%%%
\begin{figure}
\centering
\includegraphics[width=0.95\columnwidth]{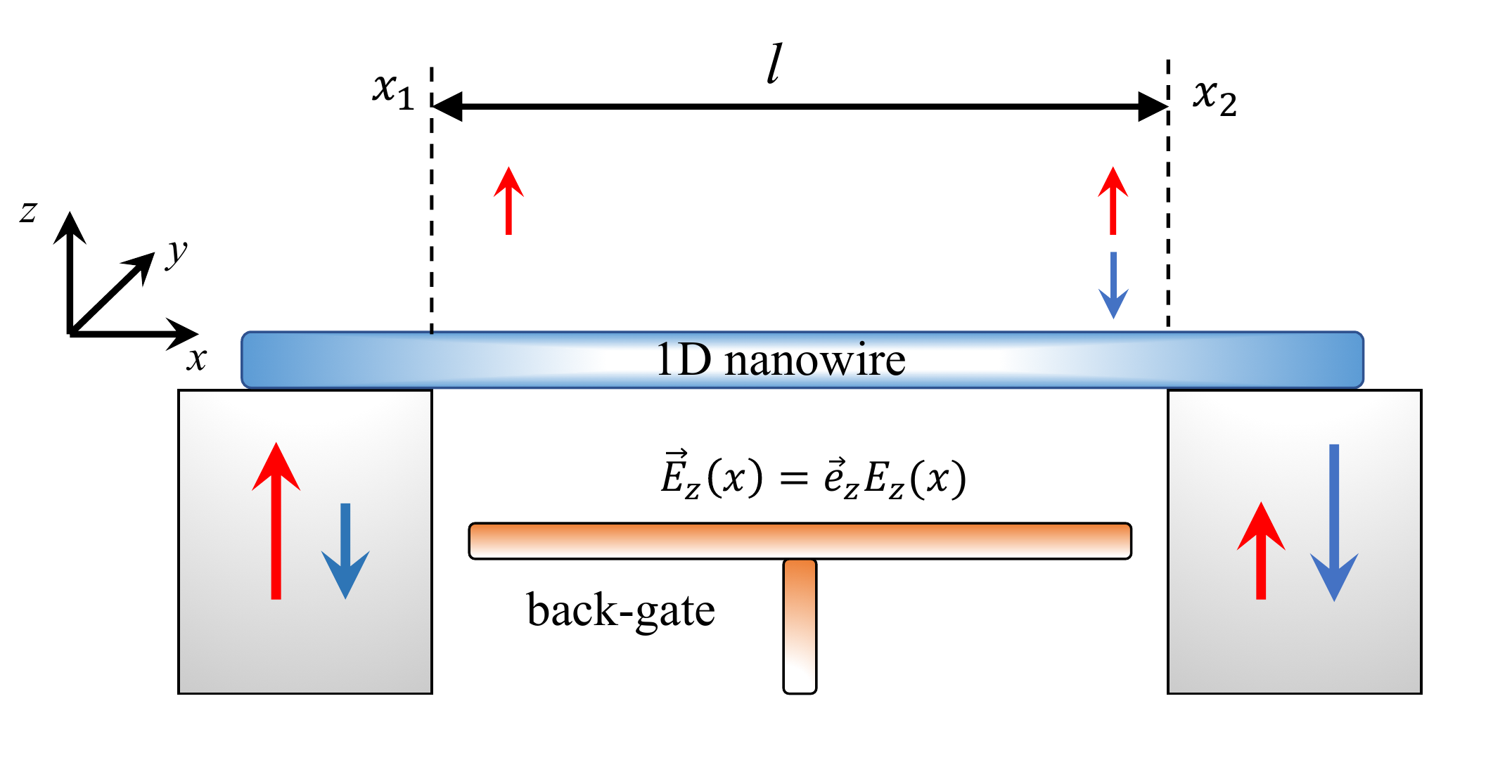}
\caption{Scheme of a nano-device. 1D nanowire of length $L$ is placed between two massive magnetically polarized electrodes. Polarization is chosen to be collinear antiparallel (AP). Spin-dependent density of states in the leads is defined through $\nu_{L\uparrow}$$=$$\nu_{R\downarrow}$$=$$\nu(1$$+$$p)$ and  $\nu_{L\downarrow}$$=$$\nu_{R\uparrow}$$=$$\nu(1$$-$$p)$. Back-gate electrode situated near the nanowire creates an electric field in $z-$ direction inducing a spin-orbit interaction (SOI) in the nanowire. Short nanowire is treated as a quantum dot (QD) (see the main text for a discussion).  We assume that QD is in a strong Coulomb blockade regime. Odd Coulomb valleys provide an access to Kondo physics. SOI leads to an accumulation of Aharonov-Casher phase in the electron wave function which is equivalent semiclassically to an electron spin precession.} \label{Fig1}
\end{figure}
%%%%%%%%%%%%%%%%%%%%%%%%%%%%%%%%%%%%%%%%%%%%%%%%%%%%%%%%%%

 One of the most remarkable manifestation of SOI in quantum devices (e.g. Datta-Das spin field-effect transistor) is associated with an accumulation of a spin-dependent phase difference in the electron (spinor) wave function [\onlinecite{ACeffect}],[\onlinecite{datta}]. This phase accumulation being controlled by an external electric field applied to a nano-device is known as Aharonov-Casher effect [\onlinecite{ACeffect}]. The electric field manipulation of the 
Aharonov-Casher interference provides a big advantage compared to external magnetic field control. In particular, no magnetization currents are generated both in nano-device and in the leads and no extra decoherence associated with an extra heating appears. Additional degree of control associated with use of ferromagnetic leads
allows to open (enhance) and close (suppress) a charge current through the nano-structure [\onlinecite{datta}],[\onlinecite{shekhterrashba}] similar to effects of a spin valve [\onlinecite{mceuen}],[\onlinecite{kobayashi}],[\onlinecite{huttel}].

In this paper we present an example when strong Coulomb blockade and established quantum coherence of electrons are 
simultaneously present and controlled in a quantum nano-wire.
We consider Kondo tunnelling of electrons through the nano-wire in the presence of strong SOI in it. We show that quantum interference originating from the Aharonov-Casher phase accumulated during tunneling process affects qualitatively many-body Kondo transmission and results in strong renormalization of Kondo temperature and significant enhancement of the charge current. 

 The paper is organized as follows: In Sec.\ref{Sec2} we introduce a model Hamiltonian of a spin-orbit active one-dimensional nanowire placed between spin-polarized electrodes and derive effective Kondo model. In Sec.\ref{Sec3} we analyse the charge current through nanowire calculated in the lowest order of perturbation theory.  In Sec.\ref{Sec4} we obtain the contribution to the charge current in second order of perturbation theory and evaluate a Kondo temperature as a function of Aharonov-Casher phase and degree of spin polarization in the leads. In Sec.\ref{Sec5} we analyze Kondo temperature in particular limit of fully polarized anti-collinear contacts.

\section{ Model}\label{Sec2} We investigate a Datta-Das spin field-effect transistor [\onlinecite{datta}] with spin-orbit active weak-link in the Kondo regime. 
We consider a one-dimensional ($1D$) nanowire embedded between two magnetically polarized electrodes in anti-parallel configuration (AP). Back gate is controlling
a spin-orbit interaction (SOI) in the nanowire, see Fig.1.
One dimensional nanowire can be treated as a quantum dot (QD) in a regime when
temperature and bias voltage are smaller compared to a mean level spacing in the QD
%charging energy 
$\delta \varepsilon$$\sim $$\hbar v_FL^{-1}$. We assume the odd number of electrons in the QD to access the Kondo regime. The model is described by the  Hamiltonian: $H$$=$$H_0$$+$$H_{tun}$, where
%%%%%%%%%%%%%%%%%%%%%%%%%%%%%%%%%
\begin{eqnarray}\label{hamiltonian}
H_0=\sum_{k,\alpha,\sigma} (\varepsilon_k-\mu_{\alpha}^{\sigma})c^{\dag}_{k\alpha\sigma}c_{k\alpha\sigma}+\sum_{\lambda}\left(\varepsilon_0d^{\dag}_{\lambda}d_{\lambda}+U_C\hat n_{\lambda}\hat n_{\bar \lambda}\right),\;\;\;\;
\end{eqnarray}
%%%%%%%%%%%%%%%%%%%%%%%%%%%%%%%%%
characterizes 1D nanowire and the magnetically polarized left (right) electrodes with chemical potentials $\mu_{L(R)}^{\sigma}$. {\color{black} Here $\varepsilon_0$ stands for the energy of first half-filled
level of the dot counted from the Fermi level of the leads, $U_C$ is the charging energy in the nanowire. The annihilation (creation) operators of the conduction electrons are denoted by $c_{k\alpha\sigma}(c^{\dag}_{k\alpha\sigma})$, where $\alpha=L,R$. The electron states in the leads are characterized by spin quantum number $\sigma=(\uparrow,\downarrow)$. The two-fold degenerate quantum level in the dot represented by linear superposition of the states with $\sigma=\uparrow$ and $\sigma=\downarrow$ is described by the pseudospin quantum number with two eigenvalues $\lambda, \bar \lambda $. We use notations $d_{\lambda}(d_{\lambda}^{\dag})$ for the electrons in the QD, $\hat n_{\lambda}=d^{\dag}_{\lambda}d_{\lambda}$, see Appendix \ref{App1}. To describe partial spin polarization of the electrodes we introduce 
spin-dependent density of states at Fermi energy: $\nu_{L\uparrow}$$=$$\nu_{R\downarrow}$$=$$\nu(1$$+$$p)$ and  $\nu_{L\downarrow}$$=$$\nu_{R\uparrow}$$=$$\nu(1$$-$$p)$, where parameter $p$ defines a degree of polarization, see Fig.\ref{Fig1}.} If magnetizations in the leads are collinear and oriented anti-parallel, the net magnetic field produced by leads  at the position of the nanowire is zero. Therefore, the net magnetic field does not lift a two-fold degeneracy of the {\color{black}pseudospin} state in the QD.  If the orientation of magnetizations is parallel, the net magnetic field at the position of the nanowire is non-zero resulting in the time-reversal symmetry breaking effects.

{\color{black} While the spin is a good quantum number in the leads, it can not be used for the characterization of the state in the nano-wire due to presence of SOI. Thus, the tunnel matrix element computed using wave functions of electrons in the leads and in QD (see, e.g., [\onlinecite{flensberg}]) is characterized by two indices $\sigma$ (spin) and $\lambda$ (pseudospin).} The most general form of the  tunneling Hamiltonian is given by
%%%%%%%%%%%%%%%%%%%%%%%%%%%%%%%%%
\begin{eqnarray}\label{tunnel}
H_{tun}=\sum_{k,\alpha;\sigma\lambda} \left(V_{k\alpha }^{\sigma\lambda}c^{\dag}_{k\alpha\sigma}d_{\lambda}+h.c\right).
\end{eqnarray}
We solve $1D$ Schr\"odinger equation for an electron in the nanowire
in the presence of SOI, see details in Appendix \ref{App1},  and express the tunnel matrix amplitudes in terms of the SOI parameters.
%One can shown \cite{footnote} that 
The two component electron wave function {\color{black}$\vec \psi_{\lambda}$} at different points are connected through the operator $\hat U$ in such way {\color{black}$\vec \psi_{\lambda}(x_2)= \hat U (x_2,x_1) \vec \psi_{\lambda}(x_1)$}, where
%%%%%%%%%%%%%%%%%%%%%%%%%%%%%%%%%
\begin{eqnarray}\label{operator}
\hat U=\exp \left[\frac{i \hat \sigma^y \vartheta (x_1,x_2)}{2}\right],\quad \vartheta=\frac{2\alpha p_F {\color{black} l} }{\hbar v_F},\;\;\;\;
\end{eqnarray}
%%%%%%%%%%%%%%%%%%%%%%%%%%%%%%%%%
characterizes accumulation of  Aharonov-Casher phase, {\color{black}$l$$=$$|x_2$$-$$x_1|$, see Fig.1.} In Eq.(\ref{operator}) $\alpha$$\propto $$E_z$ is the SOI coupling constant, $E_z$ is the electric field in $z$ direction produced by the back gate electrode, $\hat \sigma^y$ is the $y$-Pauli matrix. %\st{The Aharonov-Casher phase accumulated along the nanowire depends on the spin state.
%acts in spin Hilbert space and results in different spin up and down phase accumulation. 
%Effectively, this dependence can be interpreted as electron's spin rotation along the propagation through the area with SOI.} 
Using Eq.(\ref{operator}) and assuming that the tunneling occurs {\color{black}at the points
%at the ends of the wire 
$x_1,x_2$}, we parametrize the tunnel matrix elements as follows:
%%%%%%%%%%%%%%%%%%%%%%%%%%%%%%%%%
\begin{eqnarray}\label{parametrization}
V^{\tau\tau'}_{k  \alpha}=V_{k\alpha} \left(\delta_{\tau\tau'} \cos\left(\vartheta/4\right)\mp i\hat \sigma^y_{\tau\tau'}\sin\left(\vartheta/4\right)\right).
\end{eqnarray}
%%%%%%%%%%%%%%%%%%%%%%%%%%%%%%%%%
\noindent Here $-/+$ stands for the $L/R$ lead correspondingly. Effect of the SOI on the tunneling processes is characterized by the parameter $\vartheta$. The SOI vanishes  for $\vartheta$$=$$0$ and reaches its  maximal value at $\vartheta$$=$$\pi$, when both tunneling processes, diagonal and off-diagonal in spin (pseudospin) indices, contribute equally, see Eq.(\ref{parametrization}). We assume full  symmetry in tunneling junction {\color{black} $V_{kL}$$=$$V_{kR}$$=$$V_{tun}$}. 
%, when accumulation of the Aharonov-Casher phase leads to the two (diagonal and non-diagonal in spin indices) channels for electron tunneling (same contribution).

The mapping of the  Anderson impurity model Eq.(\ref{hamiltonian}) onto a Kondo-like model
is done using standard Schrieffer-Wolff transformation [\onlinecite{SW}], see Appendix \ref{App2}.
We assume single occupied two-fold degenerate level in the QD and 
consider the energy level width to be smaller compared to the charging energy
{\color{black}$\Gamma$$=$$2\pi\nu|V_{tun}|^2\ll U_C$}.

Effective Hamiltonian $H_{eff}$$=$$H_{dir}$$+$$H_{ex}$ contains $H_{dir}$ describing a direct (potential) electron scattering between the leads and 
\begin{eqnarray}\label{exchange}
&&H_{ex}=\sum_{\alpha}J\left[(s^z_{\alpha\alpha}S^z+s^x_{\alpha\alpha}S^x)\cos(\vartheta/2)+s^y_{\alpha\alpha}S^y\right]\nonumber\\
&&+\sum_{\alpha\neq\alpha'}J\left[s^z_{\alpha\alpha'}S^z+s^x_{\alpha\alpha'}S^x+s^y_{\alpha\alpha'}S^y\cos(\vartheta/2)\right]\nonumber\\
&&-J\sin(\vartheta/2)[ (s^z_{LL}-s^z_{RR})S^x-(s^x_{LL}-s^x_{RR})S^z]\nonumber\\
&&-\frac{J}{2}\sin(\vartheta/2) j_{LR} S^y
\end{eqnarray}
constituting the effective exchange interaction between {\color{black}pseudospin-$1/2$} in the QD, $\vec S$, and spin $\vec {s}_{\alpha\alpha'}$$=$$\sum_{kk'}(1/2)c^{\dag}_{k\alpha\sigma} {\hat \sigma}_{\sigma\sigma'}c_{k'\alpha'\sigma'}$ of the conduction electrons in the $L/R$ leads
and {\color{black}charge transfer} $j_{LR}$$=$$i\sum_{kk'}(c^{\dag}_{k L\uparrow}c_{k'R\uparrow}$$+$$c^{\dag}_{k L\downarrow}c_{k'R\downarrow} $$-$$ h.c.)$. 
We used the following notations for the exchange interaction constant in  Eq. (\ref{exchange}):
\begin{eqnarray}\label{constant}
J=2U_C\frac{|V_{tun}|^2}{|\varepsilon_0|(U_C-|\varepsilon_0|)}.
\end{eqnarray}
%Note, Hamiltonian $H_{dir}$ is proportional to another coupling constant  $H_{dir}$$\propto$$ (U$$-2|\varepsilon_0|)$. 
We concentrate below on the case of electron-hole symmetry, 
$\varepsilon_0\rightarrow(-U_C/2)$, and ignore irrelevant processes of potential scattering  \cite{footnoteM}. 

The influence of SOI effects onto the Kondo scattering has several facets. First, SOI is responsible for the different types of the spin anisotropies in the terms diagonal and off-diagonal in the lead indices (first and second lines in Eq.(\ref{exchange})). Second, SOI produces an additional coupling between the {\color{black}pseudospin} in QD and spin density of the conduction electrons aka Dzyaloshinsky-Moriya (DM) interaction ($\propto \vec e_y\cdot [\vec s_{\alpha\alpha}\times \vec S]$) [\onlinecite{pletyukhov}]. Third, SOI mediates the interaction between the {\color{black}pseudospin} in the QD and the {\color{black} charge transfer}. 

%%%%%%%%%%%%%%%%%%%%%%%%%%%%%%%%%%%%%%%%%%%%%%%%%%%%%%%%
\begin{figure}
\centering
\includegraphics[width=0.9\columnwidth]{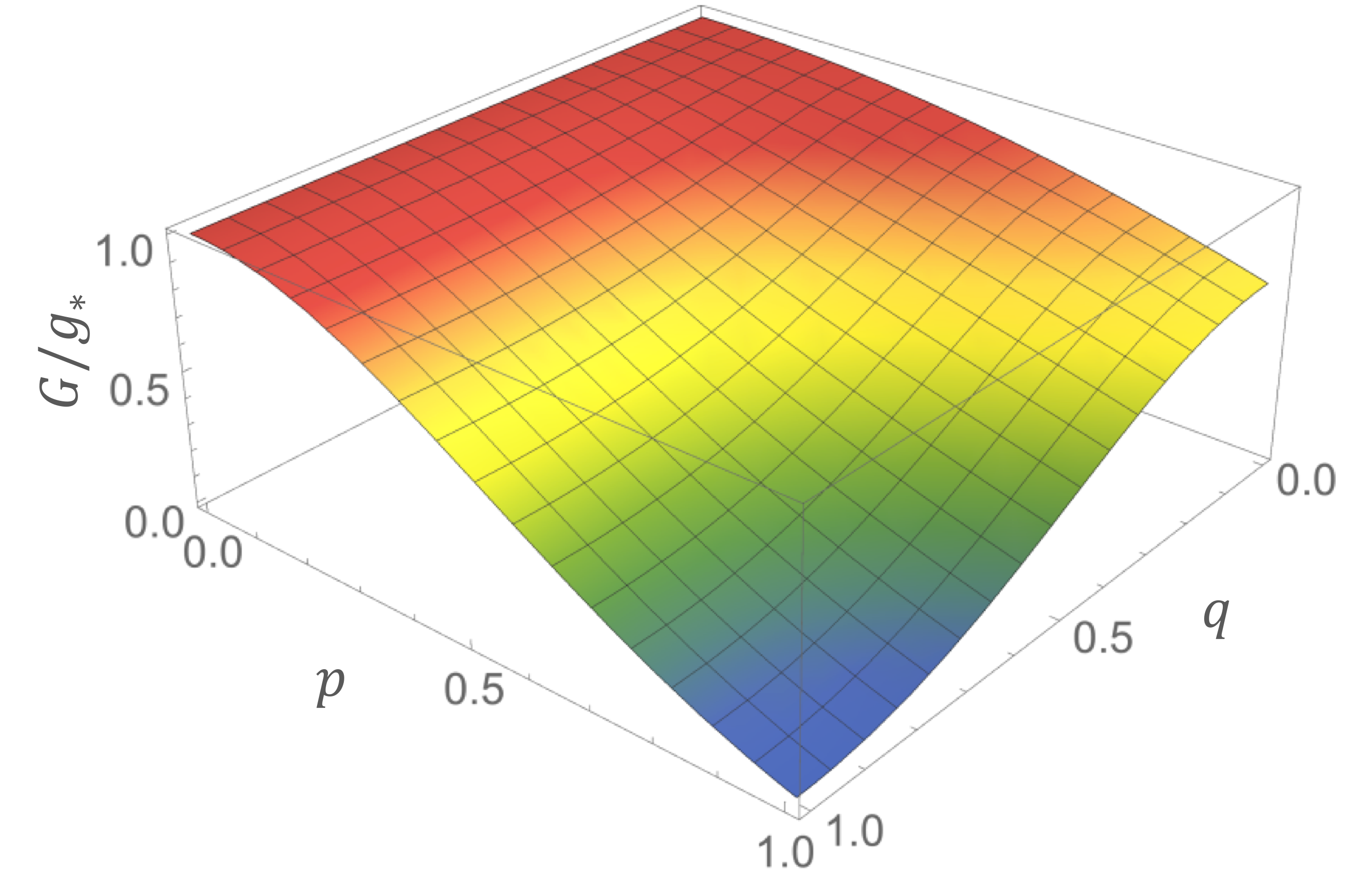}
\caption{ Differential conductance in the units of $g_{\ast}$$=$$3e^2\pi^2(\nu J)^2/(4\pi\hbar)$ as a function of both polarization in the leads $p$ and {\color{black} effects associated with the accumulation of the Aharonov-Casher phase $\vartheta$ parametrized by  $q=\cos(\vartheta/2)$ }for the case of the bias voltage $eV/T\gg 1$. } \label{Fig2}
\end{figure}
%%%%%%%%%%%%%%%%%%%%%%%%%%%%%%%%%%%%%%%%%%%%%%%%%%%%%%%%%%

\section{ Cotunneling current}\label{Sec3}
Assuming a high temperature (compared to some emerging energy scale to be defined below) regime we calculate the current through the nano-wire perturbatively in
$\nu J\ll1$.
%Let's obtain charge current through the device in the lowest (second) order in the exchange constant, 
The first non-vanishing contribution to the charge current is
$\propto (\nu J)^2$. The cotunneling current given by  Eq.(\ref{cotunneling}) can be straightforwardly derived either through
equation of motion method or using the nonequilibrium Keldysh Green function technique 
%it is easy to get from Eq.(\ref{exchange}) the next expression 
(we adopt below the units $k_B$$=$$1$):
\begin{eqnarray}\label{cotunneling}
&&I^{(2)}=\frac{e\pi^2}{4\pi\hbar}(\nu J)^2eV\left\{(1-p^2)(2-q^2)\right.\nonumber\\
&&\left.+(1+q^2)(1+p^2)-8pq\coth\left(\frac{eV}{2T}\right)\langle S^z\rangle\right\},
\end{eqnarray}
where we use shorthand notations $q$$=$$\cos(\vartheta/2)$ for parametrization of the accumulated Aharonov-Casher phase ($q$$=$$1$ for the case of the absence of  SOI,  $\vartheta=0$, and $q=0$ is when the SOI is maximal, $\vartheta=\pi$). In Eq.(\ref{cotunneling}) $\langle S^z \rangle$ denotes an out-of-equilibrium {\color{black}{\color{black} QD (nanowire) magnetization [\onlinecite{hooley}],[\onlinecite{paaske}]}}. The {\color{black}{\color{black} QD magnetization}}  that appears because of applied bias voltage in the presence of a finite polarization $p$ {\color{black}is non-vanishing} even without external magnetic field, see [\onlinecite{horing2}].  The temperature $T$ in the Eq. (\ref{cotunneling}) stands for the temperature in the contacts which are assumed to be in the equilibrium.
The expression for {\color{black} {\color{black} QD magnetization}} $\langle S^z\rangle$ is obtained from the steady-state solution of the quantum Langevin equation of motion [\onlinecite{horing2}],[\onlinecite{horing}] for the QD spin-$1/2$ in the lowest order of perturbation theory in $\nu J$: 
\begin{eqnarray}\label{magnetization}
\langle S^z\rangle=\frac{pq (eV/T)}{2(1-q^2p^2)+\varphi(\frac{eV}{T})(p^2+q^2)},
\end{eqnarray}
here $\varphi(x)$$=$$x\coth(x/2)$. 
%Note, taking into account the next order of perturbation theory yields to the forthcoming of the logarithmical correction to the out-of-equilibrium magnetization in the similar way as it appears for Kondo effect, see \cite{paaske},[Abrikosov]. However, in our work we ignore
Non-equilibrium {\color{black}{\color{black} QD magnetization}} described by Eq.(\ref{magnetization}) 
is limited by $\langle S^z \rangle $$ =$$\pm pq/(p^2+q^2)$ achieved at large bias-voltage $eV\gg T$. 
%It is the consequence of the spin-flip elastic co-tunneling processes. 

The appearance of the non-equilibrium  {\color{black}{\color{black} QD magnetization}} $\langle S^z\rangle$ in the Eq.~(\ref{cotunneling}) influences the shape of the peak in
the differential conductance, $G^{(2)}$$=$$d I^{(2)}/dV|_{V\rightarrow 0}$, see [\onlinecite{martinek2}],[\onlinecite{horing2}]. 
At large bias voltages $eV \geq T$ the effects of saturation of the {\color{black} {\color{black} QD magnetization}} result in suppression of the charge current. At low bias voltages $eV \ll T$ the contribution 
to the current proportional to the {\color{black} {\color{black} QD magnetization}} is vanishing and the current reaches 
its maximum value. The height of the conductance 
peak depends on the {\color{black}{\color{black} QD magnetization}} slope
%temperature and both, magnetic ($p$) and SOI induced ($q$), polarization as 
$\partial \langle S^z \rangle $$/$$\partial (eV)$$=$$ pq/[2T(1$$+$$p^2$$+$$q^2$$-$$p^2q^2)]$. 

The $p$- and $q$- dependence of the differential conductance given by the  Eq.(\ref{cotunneling}) is illustrated in Fig.\ref{Fig2} {\color{black} at $eV$$ \gg $$T$}. 
The leading contribution to the differential conductance calculated in the lowest non-vanishing order of the perturbation theory 
for the case of non-magnetic leads ($p$$=$$0$) in the absence of the SOI ($q$$=$$1$) 
saturates at $G^{(2)}$$=$$g_{\ast}$,  where $g_{\ast}$$=$$(e^2/\pi\hbar)(3/4)(\pi \nu J)^2$.
%differs from the unitary conductance by the value $S(S+1)\pi^2 \nu^2 J^2$).
The conductance peak at zero bias voltage for the case of partial polarization of the leads ($p\neq 1$) is $G^{(2)}$$=$$g_{\ast}(1-p^2)$.  The linear response (voltage independent) part of the differential conductance at large bias voltage ($eV$$\gg $$T$) is asymptotically given by $G^{(2)}$$=$$g_{\ast}(3-4p^2+p^4)/[3(1+p^2)]$, see [\onlinecite{martinek2}]. 
%At $p\rightarrow 0$ zero-bias anomaly dissapears.
Spin-dependent tunneling induced by SOI at $q$$<$$1$ enhances the charge transport through QD at any collinear AP of the reservoirs ($p\neq 0$). The zero-bias conductance in the case of fully polarized leads $p$$=$$1$ is given by $G^{(2)}|_{V\rightarrow0}$$=$$g_{\ast}(2/3)(1-q^2)$, while the linear response conductance
at large bias voltages is $G^{(2)}|_{V\rightarrow\infty}$$=$$g_{\ast}(2/3)(1-q^2)^2/(1+q^2)$. 
%{\color{black}(Explanation through Aharonov-Casher phase or in Intro?)}
The effect of SOI on the charge current is maximal at $q=0$. The polarization dependence of the differential conductance in this case is given by $G^{(2)}$$=$$g_{\ast}(1-p^2/3)$.

\section{ Kondo contribution to the charge current}\label{Sec4}
The next non-vanishing contribution to the charge current $\propto(\nu J)^3$ 
depends on the spin-flip processes and therefore described by the Kondo physics.
%acquirs logarithmical correction ... . 
We apply the non-equilibrium Keldysh Green's function technique and Abrikosov's {\it pseudofermion} representation [\onlinecite{abrikosov}], see details in [\onlinecite{paaske}] to proceed with the calculations. The current $I^{(3)}$$=$$I^{(3)}_K$$+$$I^{(3)}_{an}$
consists of two parts: i) $I^{(3)}_K$ is originating from the anisotropic Kondo model (first two lines in Eq.(\ref{exchange}))
%%%%%%%%%%%%%%%%%%%%%%%%%%%%%%%% 
\begin{eqnarray}\label{kondocurrent}
&&\frac{I^{(3)}_K}{g_{\ast}4\nu J}=\frac{1}{3}\left\{(1-p^2)\left[(1+q^2)V-2pq {\cal S}^z\right]+\right.\\
&&\left.q^2(1+3p^2)V-pq(1+q^2)(3+p^2) {\cal S}^z\right\} \log \left(\frac{D}{T^{\ast}}\right)\nonumber,
\end{eqnarray}
%%%%%%%%%%%%%%%%%%%%%%%%%%%%%%%%
and ii) $I^{(3)}_{an}$ is accounting for both Dzyaloshinsky-Moriya and {\color{black}charge transfer} processes in Eq.(\ref{exchange})
%%%%%%%%%%%%%%%%%%%%%%%%%%%%%%%% 
\begin{eqnarray}\label{kondocurrent2}
&&\frac{I^{(3)}_{an}}{g_{\ast}4\nu J}=\frac{ (1-q^2)}{3}(1-p^2)\left\{2V-3pq 
{\cal S}^z\right\}\log \left(\frac{D}{T^\ast}\right).\;\;\;
\end{eqnarray}
%%%%%%%%%%%%%%%%%%%%%%%%%%%%%%%%
Here $D$ is the bandwidth of the leads. We use the shorthand notations 
${\cal S}^z$$=$$V\coth(eV/2T)\langle S^z\rangle$ and $T^\ast = \max[|eV|,T]$.
%and $\log(D/|eV|)$$=$$\log(D/\sqrt{T^2+(eV)^2})$. 
The third order in $(\nu J)^3$ correction to the charge current logarithmically grows with the decreasing of both the temperature and the applied bias voltage, revealing {\it a Kondo anomaly}. The validity of perturbation theory approximation (weak coupling regime) of Eqs.(\ref{kondocurrent}),(\ref{kondocurrent2}) determines the energy scale $T_K$, the Kondo temperature. Perturbation theory breaks down at 
$T$$\lesssim $$T_K$. The effective coupling constants in this (strong coupling) regime flow {\color{black}towards} the strong coupling fixed point.
%is fulfilled ({\color{black}leading logarithmical approximation)}, where $T_K$ is {\it a Kondo temperature}.
%We obtain $T_K$ using the receipt proposed in \cite{kaminski}. 
%%%%%%%%%%%%%%%%%%%%%%%%%%%%%%%%%%%%%%%%%%%%%%%%%%%%%%%%
\begin{figure}
\centering
\includegraphics[width=0.9\columnwidth]{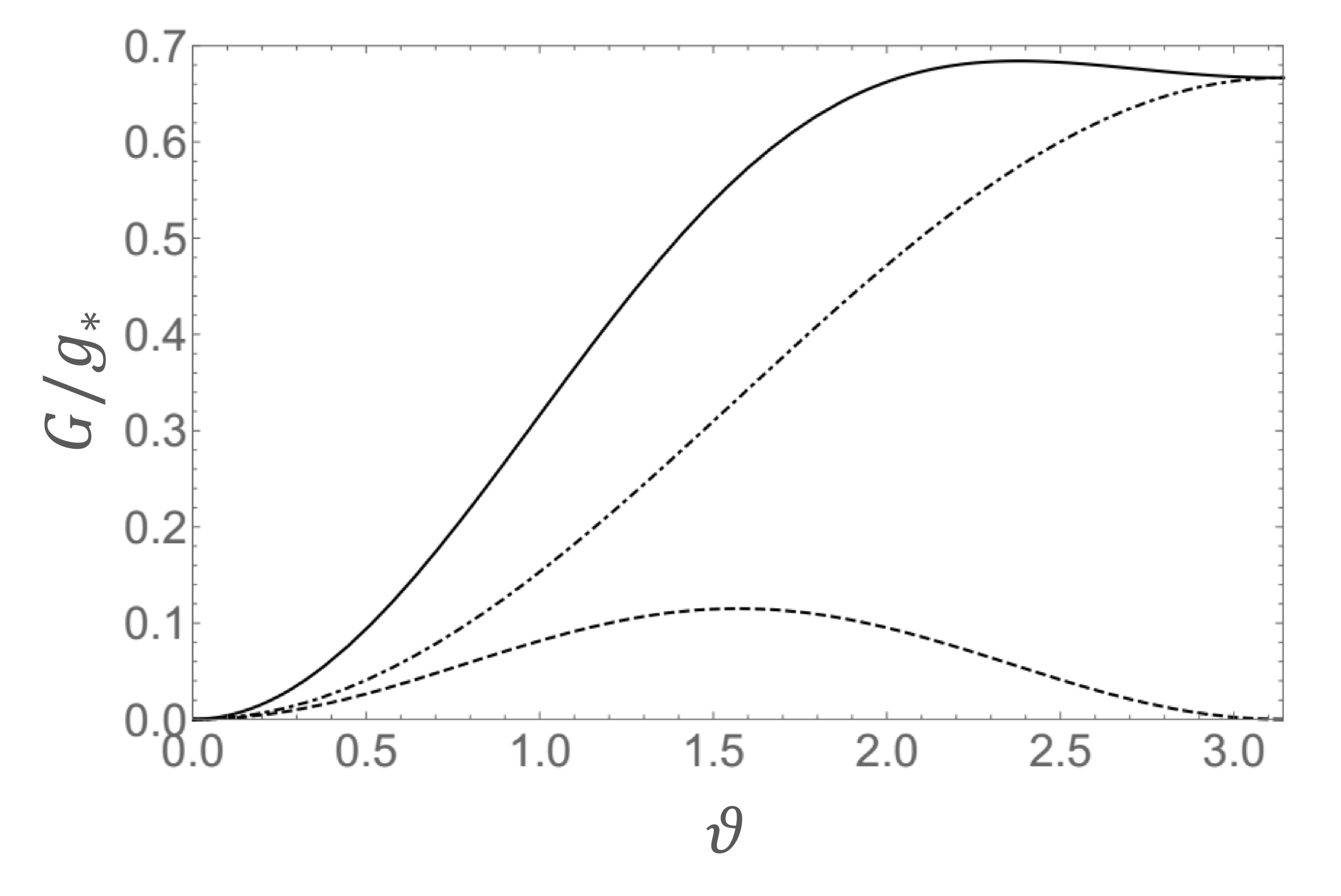}
\caption{ Dependence of zero-bias differential conductance (in units of $g_{\ast}$$=$$(3/4)(e^2/\pi\hbar)(\pi \nu J)^2$ on the Aharonov-Casher phase for the case of full AP polarization of the leads $p=1$. Dot-dashed line denotes differential conductance $G^{(2)}$ obtained from Eq.(\ref{cotunneling}), dashed line is the third order 
$\propto (\nu J)^3$ perturbative correction to the conductance $G^{(3)}$ determined by Eqs.(\ref{kondocurrent}),(\ref{kondocurrent2}). Solid line represents the sum of $G^{(2)}$ and $G^{(3)}$. Figure is plotted for the following values of model parameters: $D=1$ eV, $T=1$ K and $\nu J=0.1$. } \label{Fig3}
\end{figure}
%%%%%%%%%%%%%%%%%%%%%%%%%%%%%%%%%%%%%%%%%%%%%%%%%%%%%%%%%%
%We illustrate it by investigating the limit of the full AP polarization ($p$$=$$1$). %Considering the case of small bias voltage $eV\ll T$ and combining Eqs.(\ref{cotunneling}),(\ref{kondocurrent}) 

%Above procedure of $T_K$ estimation can be applied for any $p$ and $q$ polarization. We conclude in general Kondo temperature scales as 
The dependence of the Kondo temperature on the parameters $p$ and $q$ 
is in general determined by the solution of the system of coupled Renormalization Group (RG) equations [\onlinecite{anderson}],[\onlinecite{kaminski}]. Without losing a generality
we parametrize $T_K$ by the function $f(p,q)$ [\onlinecite{tsvelick}],[\onlinecite{andrei}]
 %%%%%%%%%%%%%%%%%%%%%%%%%%%%%%%% 
\begin{eqnarray}\label{ktemperature}
T_K = D \exp\left(-\frac{f(p,q)}{2\nu J}\right).
\end{eqnarray}
%%%%%%%%%%%%%%%%%%%%%%%%%%%%%%%%
The form of $f(p,q)$ is known for several {\color{black}limiting} cases [\onlinecite{martinek}-\onlinecite{takashi}]. In particular, $f(p,1)$$=$$1$ for all $|p|\leq 1$ [\onlinecite{martinek}],[\onlinecite{eto}]. Besides, in the case of non-magnetic leads ($p=0$) we get $f(0,q)$$=$$1$. The form of $f(p,q)$ on a line $p$$=$$1$ for $r$$=$$1$$-$$q\ll 1$ is found perturbatively from the condition of breaking down perturbation theory for the differential conductance: $f(1,r)$$=$$1$$+$$r$ (here $r$$=$$\vartheta^2/4$). With the same logic we found $f(p\ll 1, 0)$$=$$1+2p^2/3$. Perturbative analysis leads to a conclusion that $f(p,q)\geq 1$. Moreover, the absolute minimum of $f(p,q)$ is reached at symmetry lines (points), where original Hamiltonian (\ref{exchange}) is mapped onto isotropic Kondo model. We conclude that Kondo temperature in the Datta-Das transistor with AP polarized electrodes becomes function of the Aharonov-Casher phase and degree of spin polarization in the leads.

{\color{black} The most striking effect of the influence of the Aharonov-Casher interference onto Kondo scattering is manifested in the case of full AP polarization, $p$$=$$1$. In particular, the non-vanishing charge current is controlled by the Aharonov-Casher phase.}

{\color{black}The expression for the differential conductance $G$$=$$G^{(2)}$$+$$G^{(3)}$ derived in the limit of small bias voltage $eV\ll T$ 
%\st{in the case of fully AP polarized leads}
 is given by:
%%%%%%%%%%%%%%%%%%%%%%%%%%%%%%%% 
\begin{eqnarray}\label{conductance}
G= \frac{e^2}{2\pi\hbar}(\pi\nu J)^2(1-q^2)\left(1+q^24\nu J \log\frac{D}{T}\right).
\end{eqnarray}
%%%%%%%%%%%%%%%%%%%%%%%%%%%%%%%%
The conductance dependence on the Aharonov-Casher phase $\vartheta$ is shown in Fig.\ref{Fig3}. Charge current at $p$$=$$1$ and $q$$=$$1$ is blocked by the Pauli principle. However, absence of the charge current is not in a contradiction with the presence of resonant Kondo scattering. Kondo effect (and $T_K$) are fully determined by multiple {\color{black} pseudospin flip} processes
on a QD and a single (L or R) contact [\onlinecite{martinek}]. As it is seen from Eq.(\ref{conductance}), the logarithmic corrections to the conductance are positive for $J$$>$$0$ when $q\neq1$ and increase with decreasing of $T$.
%\st{Dependence of the differential conductance on Aharonov-Casher phase is shown in Fig.3.}
 While $G^{(2)}$ is monotonously increased with $\vartheta$, the behavior of $G^{(3)}$ is non-monotonous due to interplay between Kondo effect and Aharonov-Casher interferometer. As a result, the maximal current is reached at some critical value $q_{cr}$$=$$\sqrt{[1-(4\nu J)^{-1}\log(T/D)]/2}$ which depends on $T$ and initial parameters of the model.}

 \section{SOI influence on Kondo temperature}\label{Sec5}  The Hamiltonian Eq. (\ref{exchange}) casts a simple form for $p$$=$$1$
\begin{eqnarray}\label{exchange2}
\tilde{H}_{ex}=J\left\{ (s^zS^z+s^yS^y)\cos(\vartheta/2)+s^xS^x\right\}\nonumber\\
-\frac{1}{2}J\sin(\vartheta/2)\sum_{kk'}c^{\dag}_{k\gamma}c_{k'\gamma}S^x.
\end{eqnarray} 
The Hamiltonian (\ref{exchange2}) is derived from Eq. (\ref{exchange}) by
retaining the operators $c_{k\gamma}$($c_{k\gamma}^\dag$) with $\gamma$$=$$1$ for $(L,\uparrow)$, $\gamma$$=$$-1$ for $(R,\downarrow)$ and $\vec s$$=$$\sum_{kk'}(1/2)c^{\dag}_{k\gamma}\hat \sigma_{\gamma\gamma'} c_{k'\gamma'}$. We omit all other terms in Eq.(\ref{exchange}) with zero expectation values.
Eq. (\ref{exchange2}) describes an {\it anisotropic  Kondo-like model} with an additional term,  which couples the spin in the QD with the charge density in the leads. The last term in (\ref{exchange2}) can be viewed as an extra potential scattering and therefore disregarded for the particle-hole symmetric limit. 

As it is known both from the RG and exact Bethe anzatz solution of the Kondo model the maximal Kondo temperature is achieved in isotropic case [\onlinecite{tsvelick}].The Kondo temperature in anisotropic case  is
defined through the Bethe anzatz solution [\onlinecite{tsvelick}] by an equation equivalent to (\ref{ktemperature}) with the dimensionless function $f(1, q)$ dependent on the anisotropy parameter $q$$=$$\cos(\vartheta/2)$:
\begin{eqnarray}\label{ktemp2}
f(1,q)=-\frac{1}{\sqrt{1-q^2}}\log\left(\frac{1-\sqrt{1-q^2}}{q}\right).
\end{eqnarray} 
The anisotropy controlled by SOI  suppresses Kondo temperature. Asymptotic behavior of Eq.(\ref{ktemp2}) in the limit of weak ($\vartheta\rightarrow0$) SOI is given by $f(1,\vartheta)$$-$$1\approx \vartheta^2/12$. Similar behavior is also obtained from
the Renormalization Group treatment under condition $\nu J \ll 1$. While the cases of small anisotropy can be accessed perturbatively, see discussion after Eq.(\ref{ktemperature}), or by Bethe anzatz solution ($p$$=$$1$), see Eq.(\ref{ktemp2}), the solution for function $f(p,q)$ for
arbitrary values of its arguments $-1$$<$$(p,q)$$<$$1$ remains an interesting and unsolved problem \cite{parafilo}.

% The anisotropy controlled by SOI  suppresses Kondo temperature. While the cases of small anisotropy can be accessed perturbatively, see discussion after Eq.(\ref{ktemperature}), the solution for function $f(p,q)$ for
%arbitrary values of its arguments $-1$$<$$(p,q)$$<$$1$ remains an interesting and unsolved problem \cite{parafilo}. 

\section{ Summary and outlook}\label{Sec6} The interplay between resonance Kondo scattering in the quantum wire, effects of SOI in the tunnel barriers and partial spin polarization in the leads provides high level of controllability for the charge transport through the nano-device. In particular, the fine-tuning of the Kondo temperature is achieved by control of three independent tunable parameters of the system. First, the Aharonov-Casher phase is tuned by the electric field
applied to the area of the nano-wire. Second, the degree of spin-polarization in the leads is manipulated by the spin-valve [\onlinecite{mceuen}-\onlinecite{huttel}]. Third, the local out-of-equilibrium {\color{black}{\color{black} QD magnetization}} of the nano-device is controlled by the source-drain voltage. The central result of the paper 
is a prediction of a finite charge current through the nano-wire even at full AP polarization of the leads in the presence of non-zero spin-orbit interaction. {Besides, perturbative (weak coupling) calculations demonstrate pronounced (logarithmic) effects of enhancement of the current by SOI
at any given partial polarization of the leads.} Competition between the resonance scattering resulting in maximal Kondo temperature in the absence of SOI 
at $\vartheta$$=$$0$ and quantum interference due to Aharonov-Casher effect maximal at $\vartheta$$=$$\pi$ allows to find an optimal strength of SOI at $\vartheta_{cr}(q_{cr})$ under condition of  maximizing the electric current.

\vspace*{3mm}

\section*{Acknowledgements} This work was financially supported by IBS-R024-D1. AP thanks Condensed Matter and Statistical Physics section at The Abdus Salam International Centre for Theoretical Physics for the hospitality. The work of MK was  performed  in  part  at  Aspen Center for Physics, which is supported by National Science  Foundation  grant  PHY-1607611 and partially supported by a grant from the Simons Foundation.

\begin{appendix}
%\begin{widetext}
\setcounter{equation}{0}
\setcounter{figure}{0}
\setcounter{table}{0}
\makeatletter
\renewcommand{\theequation}{S\arabic{equation}}
\renewcommand{\thefigure}{S\arabic{figure}}
\renewcommand{\bibnumfmt}[1]{[S#1]}
\renewcommand{\citenumfont}[1]{S#1}
\numberwithin{equation}{section}

\vspace*{3mm}

\section{Spin-orbit interaction and spin-dependent tunnel matrix elements}\label{App1}
In this Appendix we calculate the Aharonov-Casher phase $\vartheta$, see Eq.(\ref{operator}). We start from a Schr\"odinger equation for the electron in $1D$ nanowire in the presence of external homogeneous electric field $E_z$ produced by the back-gate and directed perpendicular to nanowire:
%%%%%%%%%%%%%%%%%%%%%
\begin{eqnarray}\label{Schr}
-\frac{\hbar^2}{2m}\frac{\partial^2 \vec\psi}{\partial x^2}-\alpha\hat \sigma^y i\hbar\frac{\partial \vec\psi}{\partial x}=(E_F-\varepsilon) \vec \psi,
\end{eqnarray}
%%%%%%%%%%%%%%%%%%%%%%
where $\vec \psi$ is a two component electron wave function (spinor), $\alpha\propto E_z$ is a spin-orbit interaction coupling constant. Since $\varepsilon \leq \hbar v_F L^{-1}\ll E_F$, we can present the electron's wave function in terms of right and left moving parts:
%%%%%%%%%%%%%%%%%%%%%
\begin{eqnarray}\label{wf}
\vec \psi=e^{ip_Fx/\hbar}\vec \psi_{+}(x)+e^{-ip_Fx/\hbar}\vec \psi_{-}(x),
\end{eqnarray}
%%%%%%%%%%%%%%%%%%%%%%
where $p_F=\sqrt{2m(E_F-\varepsilon)}$ is a Fermi momentum. Substituting the wave function Eq.(\ref{wf}) into Eq.(\ref{Schr}) and neglecting the second derivative we get:
%%%%%%%%%%%%%%%%%%%%%
\begin{eqnarray}\label{Schr2}
- iv_F\hbar \frac{\partial \vec \psi_{\pm}}{\partial x}= \alpha p_F\hat \sigma^y \vec \psi_{\pm}.
\end{eqnarray}
%%%%%%%%%%%%%%%%%%%%%%
From this equation one can see that spinor function $\vec \psi_{\pm}$ satisfy relation
%%%%%%%%%%%%%%%%%%%%%
\begin{eqnarray}\label{wf2}
\vec \psi_{\pm}(x)= \hat U(x)\vec \psi_{\pm}(0)\quad,\quad \hat U(x)=e^{i\hat\sigma^y\vartheta(x)/2},
\end{eqnarray}
%%%%%%%%%%%%%%%%%%%%%%
where $\vartheta(x)=2\alpha x p_F/(\hbar v_F)$ is an Aharonov-Casher phase. 

Despite the fact that in the presence of SOI electronic spin is unsuitable quantum number for the classification of the electronic states, the energy levels continue to be double degenerate. If $\vec \psi_{\lambda}$ is the eigenstate with energy $\varepsilon$, then $\vec \psi_{\bar \lambda}=i\hat \sigma^y \vec \psi_{\lambda}^{\ast}$ is also eigenstate with the same energy. Here we used notations $\vec \psi_{\lambda(\bar \lambda)}$ for the normalized electron wave function in the nanowire. As this takes place one can classified the states by the spin structure of the wave functions at fixed point, for example, at $x=0$ middle point of the nanowire between left and right electrodes. Assuming that eigenstates $\vec \psi_{\lambda(\bar \lambda)}(0)$ in the middle of nanowire correspond to the state with spin up and spin down, we define the value of the wave function in the point $x_1, x_2$ where tunneling into the leads occurs ($x_{1(2)}=\pm l/2$):
\begin{widetext}
%%%%%%%%%%%%%%%%%%%%%
\begin{eqnarray}\label{wftunn}
&&\vec \psi_{\lambda}(x_1)=\hat U(x_1, 0)\left(
\begin{array}{c}
1 \\ 0
\end{array}\right)=\left(
\begin{array}{c}
\cos(\vartheta/4) \\ \sin(\vartheta/4)
\end{array}\right)\quad,\quad \vec \psi_{\lambda}(x_2)=\hat U(x_2, 0)\left(
\begin{array}{c}
1 \\ 0
\end{array}\right)=\left(
\begin{array}{c}
\cos(\vartheta/4) \\ -\sin(\vartheta/4)
\end{array}\right),\\
&&\vec \psi_{\bar \lambda}(x_1)=\hat U(x_1, 0)\left(
\begin{array}{c}
0 \\ 1
\end{array}\right)=\left(
\begin{array}{c}
-\sin(\vartheta/4) \\ \cos(\vartheta/4)
\end{array}\right)\quad,\quad \vec \psi_{\bar \lambda}(x_2)=\hat U(x_2, 0)\left(
\begin{array}{c}
0 \\ 1
\end{array}\right)=\left(
\begin{array}{c}
\sin(\vartheta/4) \\ \cos(\vartheta/4)
\end{array}\right).
\end{eqnarray}
%%%%%%%%%%%%%%%%%%%%%%
\end{widetext}
Amplitude of electron tunneling from the left lead to the nanowire in state $\lambda$, $\bar \lambda$ can be found as follows
\begin{widetext}
%%%%%%%%%%%%%%%%%%%%%
\begin{eqnarray}\label{tunnmatrixelem}
&&V_{kL}^{\uparrow\lambda}=\sum_k V_{kL}\langle (\uparrow, 0)|\vec \psi_{\lambda}(x_1)\rangle \quad,\quad 
V_{kL}^{\downarrow \lambda}=\sum_k V_{kL}\langle (0, \downarrow)|\vec \psi_{\lambda}(x_1)\rangle, \\
&&V_{kL}^{\uparrow\bar\lambda}=\sum_k V_{kL}\langle (\uparrow, 0)|\vec \psi_{\bar\lambda}(x_1)\rangle \quad,\quad 
V_{kL}^{\downarrow \bar\lambda}=\sum_k V_{kL}\langle (0, \downarrow)|\vec \psi_{\bar\lambda}(x_1)\rangle,
\end{eqnarray}
%%%%%%%%%%%%%%%%%%%%%%
\end{widetext}
where $V_{kL}$ is the transition amplitude. The tunnel matrix element for tunneling processes from the right lead can be defined in similar way by using wave functions $\vec \psi_{\lambda(\bar \lambda)}(x_2)$. 

\section{Schrieffer-Wolff transformation}\label{App2}

In this section we derive effective Kondo Hamiltonian for the general case of spin-dependent tunnel matrix elements. The mapping of the Anderson-like  impurity model Eqs.(\ref{hamiltonian}-\ref{tunnel}) onto a Kondo-like model is done using a standard Schrieffer-Wolff transformation:
%%%%%%%%%%%%%%%%%%%%%
\begin{eqnarray}\label{SW}
H_{eff}=e^SHe^{-S}\equiv H+[S,H]+\frac{1}{2}[S,[S,H]]+... .
\end{eqnarray}
%%%%%%%%%%%%%%%%%%%%%%
The first step is to eliminate the first order in tunneling amplitude terms  using the following condition:
%%%%%%%%%%%%%%%%%%%%%
\begin{eqnarray}\label{cond}
[S, H_0]=-H_{tun},
\end{eqnarray}
%%%%%%%%%%%%%%%%%%%%%%
As a result, the effective Hamiltonian is transformed to
%%%%%%%%%%%%%%%%%%%%%
\begin{eqnarray}\label{K}
H_{eff}=H_0+\frac{1}{2}[S,H_{tun}]+... .
\end{eqnarray}
%%%%%%%%%%%%%%%%%%%%%%
We choose operator $S$ in the following form:
%%%%%%%%%%%%%%%%%%%%%
\begin{eqnarray}\label{S}
S=\left[\sum (A_{\alpha\sigma\lambda}+B_{\alpha\sigma\lambda}\hat n_{\bar \lambda})c^{\dag}_{k\alpha\sigma}d_\lambda -h.c. \right].
\end{eqnarray}
%%%%%%%%%%%%%%%%%%%%%%
Using condition given by Eq.(\ref{cond}) one can determine the constants $A_{\alpha\sigma\lambda}$, $B_{\alpha\sigma\lambda}$.
After straightforward calculations using Eqs.(\ref{K}-\ref{S}) we obtain the effective Hamiltonian $H_{eff}=H_{dir}+H_{ex}$. The first term responsible for the electron potential scattering between leads is given by
\begin{widetext}
%%%%%%%%%%%%%%%%%%%%%
\begin{eqnarray}\label{direct}
H_{dir}=\sum \frac{K_{\alpha\alpha'}}{4}\left[ \frac{n_{\alpha\alpha'}}{2}\left(V_{k\alpha}^{\uparrow\uparrow}(V_{k'\alpha'}^{\uparrow\uparrow})^{\ast}+V_{k\alpha}^{\downarrow\downarrow}(V_{k'\alpha'}^{\downarrow\downarrow})^{\ast}+V_{k\alpha}^{\downarrow\uparrow}(V_{k'\alpha'}^{\uparrow\downarrow})^{\ast}+V_{k\alpha}^{\uparrow\downarrow}(V_{k'\alpha'}^{\downarrow\uparrow})^{\ast}\right)\right.\nonumber\\
+ s^z_{\alpha\alpha'}\left (V_{k\alpha}^{\uparrow\uparrow}(V_{k'\alpha'}^{\uparrow\uparrow})^{\ast}-V_{k\alpha}^{\downarrow\downarrow}(V_{k'\alpha'}^{\downarrow\downarrow})^{\ast}-V_{k\alpha}^{\downarrow\uparrow}(V_{k'\alpha'}^{\uparrow\downarrow})^{\ast}+V_{k\alpha}^{\uparrow\downarrow}(V_{k'\alpha'}^{\downarrow\uparrow})^{\ast}\right)\nonumber\\
+ s^x_{\alpha\alpha'} \left(V_{k\alpha}^{\uparrow\uparrow}(V_{k'\alpha'}^{\uparrow\downarrow})^{\ast}
+V_{k\alpha}^{\uparrow\downarrow}(V_{k'\alpha'}^{\downarrow\downarrow})^{\ast}+V_{k\alpha}^{\downarrow\downarrow}(V_{k'\alpha'}^{\downarrow\uparrow})^{\ast}+V_{k\alpha}^{\downarrow\uparrow}(V_{k'\alpha'}^{\uparrow\uparrow})^{\ast}\right)\nonumber\\
\left.+i s^y_{\alpha\alpha'} \left(V_{k\alpha}^{\uparrow\uparrow}(V_{k'\alpha'}^{\uparrow\downarrow})^{\ast}
+V_{k\alpha}^{\uparrow\downarrow}(V_{k'\alpha'}^{\downarrow\downarrow})^{\ast}-V_{k\alpha}^{\downarrow\downarrow}(V_{k'\alpha'}^{\downarrow\uparrow})^{\ast}-V_{k\alpha}^{\downarrow\uparrow}(V_{k'\alpha'}^{\uparrow\uparrow})^{\ast}\right)\right].
\end{eqnarray}
%%%%%%%%%%%%%%%%%%%%%%
The matrix elements used in Eq.(\ref{direct}) are given by %%%%%%%%%%%%%%%%%%%%%
\begin{eqnarray}\label{ex1}
K_{\alpha\alpha'}=\frac{1}{\varepsilon_{k\alpha}-\varepsilon_0}+\frac{1}{\varepsilon_{k'\alpha'}-\varepsilon_0}-\frac{1}{U_C+\varepsilon_0-\varepsilon_{k\alpha}}-\frac{1}{U_C+\varepsilon_0-\varepsilon_{k'\alpha'}},
\end{eqnarray}
%%%%%%%%%%%%%%%%%%%%%%
%Note, that the effective Hamiltonian (\ref{direct}) was calculated in \cite{shekhter} for the case of resonant level (unoccupied energy level, $\varepsilon_0>0$).

The next step is to find the Hamiltonian responsible for the exchange processes:
%%%%%%%%%%%%%%%%%%%%%
\begin{eqnarray}\label{exchange3}
H_{ex}=H_{1}+H_{2}+H_{3},
\end{eqnarray}
%%%%%%%%%%%%%%%%%%%%%%
where the Hamiltonian $H_1$ 
%%%%%%%%%%%%%%%%%%%%%
\begin{eqnarray}\label{H}
H_{1}=\sum \frac{J_{\alpha\alpha'}}{2}\left[ s^z_{\alpha\alpha'}  S^z \left(V_{k\alpha}^{\uparrow\uparrow}(V_{k'\alpha'}^{\uparrow\uparrow})^{\ast}+V_{k\alpha}^{\downarrow\downarrow}(V_{k'\alpha'}^{\downarrow\downarrow})^{\ast}-V_{k\alpha}^{\downarrow\uparrow}(V_{k'\alpha'}^{\uparrow\downarrow})^{\ast}-V_{k\alpha}^{\uparrow\downarrow}(V_{k'\alpha'}^{\downarrow\uparrow})^{\ast}\right)\right.\nonumber\\
+ s^x_{\alpha\alpha'}  S^x \left(V_{k\alpha}^{\uparrow\uparrow}(V_{k'\alpha'}^{\downarrow\downarrow})^{\ast}+V_{k\alpha}^{\downarrow\downarrow}(V_{k'\alpha'}^{\uparrow\uparrow})^{\ast}+V_{k\alpha}^{\downarrow\uparrow}(V_{k'\alpha'}^{\downarrow\uparrow})^{\ast}+V_{k\alpha}^{\uparrow\downarrow}(V_{k'\alpha'}^{\uparrow\downarrow})^{\ast}\right)\nonumber\\
+ s^y_{\alpha\alpha'}  S^y \left(V_{k\alpha}^{\uparrow\uparrow}(V_{k'\alpha'}^{\downarrow\downarrow})^{\ast}+V_{k\alpha}^{\downarrow\downarrow}(V_{k'\alpha'}^{\uparrow\uparrow})^{\ast}-V_{k\alpha}^{\downarrow\uparrow}(V_{k'\alpha'}^{\downarrow\uparrow})^{\ast}-V_{k\alpha}^{\uparrow\downarrow}(V_{k'\alpha'}^{\uparrow\downarrow})^{\ast}\right),
\end{eqnarray}
%%%%%%%%%%%%%%%%%%%%%%
is equivalent to an anisotropic Kondo model. The Hamiltonian $H_2$ 
%%%%%%%%%%%%%%%%%%%%%
\begin{eqnarray}\label{H2}
H_{2}=\sum \frac{J_{\alpha\alpha'}}{4}\left[n_{\alpha\alpha'} S^z\left(V_{k\alpha}^{\uparrow\uparrow}(V_{k'\alpha'}^{\uparrow\uparrow})^{\ast}-V_{k\alpha}^{\downarrow\downarrow}(V_{k'\alpha'}^{\downarrow\downarrow})^{\ast}+
V_{k\alpha}^{\downarrow\uparrow}(V_{k'\alpha'}^{\uparrow\downarrow})^{\ast}-V_{k\alpha}^{\uparrow\downarrow}(V_{k'\alpha'}^{\downarrow\uparrow})^{\ast}\right)\right.\nonumber\\ 
+n_{\alpha\alpha'} S^x\left(V_{k\alpha}^{\uparrow\uparrow}(V_{k'\alpha'}^{\downarrow\uparrow})^{\ast}
+V_{k\alpha}^{\uparrow\downarrow}(V_{k'\alpha'}^{\uparrow\uparrow})^{\ast}+V_{k\alpha}^{\downarrow\downarrow}(V_{k'\alpha'}^{\uparrow\downarrow})^{\ast}+V_{k\alpha}^{\downarrow\uparrow}(V_{k'\alpha'}^{\downarrow\downarrow})^{\ast}\right)\nonumber\\
\left.+in_{\alpha\alpha'} S^y\left(-V_{k\alpha}^{\uparrow\uparrow}(V_{k'\alpha'}^{\downarrow\uparrow})^{\ast}
+V_{k\alpha}^{\uparrow\downarrow}(V_{k'\alpha'}^{\uparrow\uparrow})^{\ast}+V_{k\alpha}^{\downarrow\downarrow}(V_{k'\alpha'}^{\uparrow\downarrow})^{\ast}-V_{k\alpha}^{\downarrow\uparrow}(V_{k'\alpha'}^{\downarrow\downarrow})^{\ast}\right)\right]
\end{eqnarray}
%%%%%%%%%%%%%%%%%%%%%%
describes the coupling between the spin-$1/2$ on the dot and the "charge" density in the leads $n_{\alpha\alpha'}$. 

The Hamiltonian $H_3$
%%%%%%%%%%%%%%%%%%%%%
\begin{eqnarray}\label{H3}
H_{3}=\sum \frac{J_{\alpha\alpha'}}{2}\left[i s^y_{\alpha\alpha'} S^x\left(V_{k\alpha}^{\uparrow\uparrow}(V_{k'\alpha'}^{\downarrow\downarrow})^{\ast}-V_{k\alpha}^{\downarrow\downarrow}(V_{k'\alpha'}^{\uparrow\uparrow})^{\ast}-V_{k\alpha}^{\downarrow\uparrow}(V_{k'\alpha'}^{\downarrow\uparrow})^{\ast}+V_{k\alpha}^{\uparrow\downarrow}(V_{k'\alpha'}^{\uparrow\downarrow})^{\ast}\right)\right.\nonumber\\
+i s^x_{\alpha\alpha'}S^y\left(-V_{k\alpha}^{\uparrow\uparrow}(V_{k'\alpha'}^{\downarrow\downarrow})^{\ast}+V_{k\alpha}^{\downarrow\downarrow}(V_{k'\alpha'}^{\uparrow\uparrow})^{\ast}-V_{k\alpha}^{\downarrow\uparrow}(V_{k'\alpha'}^{\downarrow\uparrow})^{\ast}+V_{k\alpha}^{\uparrow\downarrow}(V_{k'\alpha'}^{\uparrow\downarrow})^{\ast}\right)\nonumber\\
+ s^x_{\alpha\alpha'} S^z\left(V_{k\alpha}^{\uparrow\uparrow}(V_{k'\alpha'}^{\uparrow\downarrow})^{\ast}
-V_{k\alpha}^{\uparrow\downarrow}(V_{k'\alpha'}^{\downarrow\downarrow})^{\ast}-V_{k\alpha}^{\downarrow\downarrow}(V_{k'\alpha'}^{\downarrow\uparrow})^{\ast}+V_{k\alpha}^{\downarrow\uparrow}(V_{k'\alpha'}^{\uparrow\uparrow})^{\ast}\right)\nonumber\\
+ s^z_{\alpha\alpha'} S^x
\left(V_{k\alpha}^{\uparrow\uparrow}(V_{k'\alpha'}^{\downarrow\uparrow})^{\ast}
+V_{k\alpha}^{\uparrow\downarrow}(V_{k'\alpha'}^{\uparrow\uparrow})^{\ast}-V_{k\alpha}^{\downarrow\downarrow}(V_{k'\alpha'}^{\uparrow\downarrow})^{\ast}-V_{k\alpha}^{\downarrow\uparrow}(V_{k'\alpha'}^{\downarrow\downarrow})^{\ast}\right)\nonumber\\
+is^y_{\alpha\alpha'} S^z\left(
V_{k\alpha}^{\uparrow\uparrow}(V_{k'\alpha'}^{\uparrow\downarrow})^{\ast}
-V_{k\alpha}^{\uparrow\downarrow}(V_{k'\alpha'}^{\downarrow\downarrow})^{\ast}
+V_{k\alpha}^{\downarrow\downarrow}(V_{k'\alpha'}^{\downarrow\uparrow})^{\ast}
-V_{k\alpha}^{\downarrow\uparrow}(V_{k'\alpha'}^{\uparrow\uparrow})^{\ast}\right)\nonumber\\
\left.+i s^z_{\alpha\alpha'} S^y\left(-V_{k\alpha}^{\uparrow\uparrow}(V_{k'\alpha'}^{\downarrow\uparrow})^{\ast}
+V_{k\alpha}^{\uparrow\downarrow}(V_{k'\alpha'}^{\uparrow\uparrow})^{\ast}
-V_{k\alpha}^{\downarrow\downarrow}(V_{k'\alpha'}^{\uparrow\downarrow})^{\ast}
+V_{k\alpha}^{\downarrow\uparrow}(V_{k'\alpha'}^{\downarrow\downarrow})^{\ast}\right)\right],
\end{eqnarray}
%%%%%%%%%%%%%%%%%%%%%%
accounts for the Dzyaloshinsky-Moriya-like interaction.

Exchange coupling constant reads as follows 
%%%%%%%%%%%%%%%%%%%%%
\begin{eqnarray}\label{ex2}
J_{\alpha\alpha'}=\frac{1}{\varepsilon_{k\alpha}-\varepsilon_0}+\frac{1}{U_C+\varepsilon_0-\varepsilon_{k\alpha}}+\frac{1}{\varepsilon_{k'\alpha'}-\varepsilon_0}+\frac{1}{U_C+\varepsilon_0-\varepsilon_{k'\alpha'}}.
\end{eqnarray}
\end{widetext}
%%%%%%%%%%%%%%%%%%%%%%

Substituting parametrization Eq.(\ref{parametrization}) into Eqs.(\ref{ex1})-(\ref{exchange3}) and considering exchange coupling constants of conduction electrons at Fermi energy, $\varepsilon_{k\alpha}=\varepsilon_k-\mu_{\alpha}^{\sigma}\approx 0$, we can obtain Hamiltonian Eq.(\ref{exchange}).
%%%%%%%%%%%%%%%%%%%%%%
%\end{widetext}

\end{appendix}

%\vspace*{-5mm}


\begin{thebibliography}{99}
\bibitem{footnoteM}
 The Hamiltonian (\ref{exchange}) acquires more complicated form in case of asymmetry induced by tunnel coupling ($V_{kL}\neq V_{kR}$) or by different polarization in $L/R$ leads ($\nu_{L\uparrow(\downarrow)}\neq \nu_{R\downarrow(\uparrow)}$), see Eqs.(\ref{H}-\ref{H3}). The most relevant additional term in this case is an effective Zeeman interaction, $\propto \sum_{\alpha}J_{\alpha\alpha}(\nu_{\alpha\uparrow}-\nu_{\alpha\downarrow})S^z$ (where $J_{\alpha\alpha}\propto |V_{k\alpha}|^2$ is exchange constant). It results in lifting of two-fold degeneracy of the pseudospin state in the QD. The effects of non-collinear polarization of $L/R$ leads are beyond the scope of this paper.
\end{thebibliography}
\end{document}